\documentclass[prl,twocolumn,showpacs,floatfix]{revtex4}

\usepackage{graphicx}
\usepackage{amsmath}

\begin{document}

\title{Simple proof of equivalence between adiabatic quantum computation and
the circuit model}
\author{Ari Mizel$^1$, Daniel A. Lidar$^2$, and Morgan Mitchell$^3$}
\affiliation{$^1$Department of Physics, Pennsylvania State University, University Park, Pennsylvania 16802, U.S.A.}
\affiliation{$^2$Departments of Chemistry, Electrical Engineering-Systems, and
Physics, University of Southern California, Los Angeles, California 90089,
U.S.A.}
\affiliation{$^3$Institut de Ciències Fotòniques 08860 Castelldefels, (Barcelona), Spain}

\begin{abstract}
We prove the equivalence between adiabatic quantum computation and quantum
computation in the circuit model. An explicit adiabatic computation
procedure is given that generates a ground state from which the answer can
be extracted. The amount of time needed is evaluated by computing the gap.
We show that the procedure is computationally efficient.
\end{abstract}

\pacs{03.67.Lx}
\maketitle

\textit{Introduction.}--- In the effort to realize a quantum computer (QC),
adiabatic quantum computation \cite{Farhi:00} offers a promising alternative
to the standard \textquotedblleft circuit model\textquotedblright\ \cite%
{Deutsch:89,DiVincenzo:00}. In comparison to the circuit model, AQC\
alleviates the need to perform fast quantum logic operations and
measurements, which is particularly troublesome in the context of fault
tolerant quantum computation \cite{AlickiLidarZanardi:05}. In AQC, the
answer to a calculation is contained in the ground state of a quantum
Hamiltonian. By placing a system in the ground state of a simple Hamiltonian
and then adiabatically changing until the desired Hamiltonian is reached,
one carries the system into the computationally meaningful state. The AQC
model was known from the outset to be efficiently simulatable by the
standard model \cite{Farhi:00,vanDam:01}, but for some time researchers
wondered whether AQC\ could efficiently simulate the standard model.

Recently, a number of relatively complex proofs of the equivalence between
the circuit model and AQC\ were given. One proof showed that AQC\ using
Hamiltonians with long-range five- or three-body interactions, or
nearest-neighbor two-body interactions with six-state particles, can
efficiently simulate the circuit model \cite{Aharonov:04}. This result was
soon modified to qubits with two-body interactions \cite{Kempe:04,Siu:04},
and then it was shown that AQC using qubits with nearest-neighbor two-body
interactions on a 2D lattice can efficiently simulate the standard circuit
model \cite{Oliveira:05}. The proofs in Refs. \cite%
{Aharonov:04,Kempe:04,Siu:04,Oliveira:05} all start from a five-body
interaction Hamiltonian that arises in Kitaev's quantum NP-complete
\textquotedblleft local Hamiltonians\textquotedblright\ problem \cite%
{Kitaev:book}. They then require a reduction to two-body Hamiltonians, and a
proof that the spectral gap of the AQC Hamiltonian thus constructed is
properly lower bounded. Let the number of algorithm steps
(number of single- and two-qubit gates)
be $N$. The running time is $O(N^{5})$ with five-body
interactions and $O(N^{14})$ with three-body interactions \cite{Aharonov:04}%
, which was improved to $O(N^{12})$ with two-body interactions in Ref. \cite%
{Siu:04}. An additional improvement by a factor of $N$ was given in Ref. 
\cite{Deift:06}, where relatively simple methods were used to provide a
lower bound on the minimal energy gap.

Here we provide an alternative, constructive proof of the equivalence
between the standard circuit model and AQC that is physically and
mathematically transparent, amenable to implementation and yields a running
time $T$ of order $(MN)^{2}$ or better, where $M$ is the number of
qubits.
E.g., in the case of Shor's algorithm for factoring an $L$-bit
integer using a linear nearest-neighbor
qubit array \cite{Fowler:04}, this translates into $T\sim [(2L+4)(8L^4)]^2 \sim 256
L^{10}$ compared to $T \sim (8L^4)^{11}\sim 10^{10}L^{44}$ using the
previous $O(N^{11})$ scaling. We do this
by setting up an explicit Hamiltonian involving at most two-body,
nearest-neighbor interactions between particles on a 2D\ lattice. Our
construction uses the method of ground state quantum computation (GSQC),
which was independently proposed in Refs. \cite{Mizel:01,Mizel:02,Mizel:04}
around the same time as AQC and also studied in Ref.~\cite{Mao:05}.

In GSQC, one executes an algorithm by producing a ground state that \emph{%
spatially} encodes the entire temporal trajectory of the algorithm, from
input to output. This requires $N$ times as much hardware but provides some
robustness against decoherence. GSQC\ was deliberately constructed to
simulate the standard model \cite{Mizel:01}. However, little attention was
devoted to the process of reaching the desired ground state. Here, we marry
together AQC and GSQC. The result is a formalism supplying an explicit
Hamiltonian $H(s)$ acting on qubits with at most two-body nearest neighbor
interactions for any algorithm formulated in the circuit model. The initial
Hamiltonian $H(0)$ and its ground state are simple. The intermediate
Hamiltonian $H(s)$ ($0\leq s\leq 1$) has a gap and a non-degenerate ground
state for all $s$ (the dimensionless time). The final Hamiltonian $H(1)$ has
a ground state containing the solution to the algorithm. Using the adiabatic
theorem, we provide an upper bound on the time needed to reach $H(1)$ while
keeping the system in its ground state. This bound scales polynomially in
the number of algorithm steps and qubits; the calculation is efficient.

\textit{Single qubit.}--- 
The ground state that contains the result of a given standard algorithm is
specified as follows \cite{Mizel:01}. First consider a particularly simple
computation involving only a single qubit with basis states $|0\rangle $ and $%
|1\rangle $. In the circuit model the qubit evolves through $N+1$ time
steps: its initial state and a state after each algorithm step.  If
the initial state is $|0\rangle $ and algorithm step $i$ consists of
application of a $2\times 2$ unitary gate $U_{i}$, then the two
amplitudes at time step $i$ are given by $U_{i}\cdots U_{1}|0\rangle
$, where $1\leq i\leq N$. Since there are $2$ amplitudes at each of the $N+1$ steps, the whole
trajectory can be described by giving $2(N+1)$ complex amplitudes. In GSQC, instead of a
time-dependent state in a $2$-dimensional Hilbert space, the qubit has a
time-independent state in a $2(N+1)$-dimensional Hilbert space, with basis
states $c_{i,0}^{\dagger }\left\vert \mathrm{vac}\right\rangle $ and $%
c_{i,1}^{\dagger }\left\vert \mathrm{vac}\right\rangle $, $i=0,\dots ,N$.
Here, $c_{i,x}^{\dagger }$ ($c_{i,x}$) is a fermionic creation (annihilation)
operator for a particle in state $x\in \{0,1\}$ in mode $i\in
\{0,1,...N\}$ \cite{comment}.   The amplitude of the time-independent
wavefunction in basis state $c_{i,0}^{\dagger }\left\vert
\mathrm{vac}\right\rangle $ ($c_{i,1}^{\dagger }\left\vert \mathrm{vac}\right\rangle $) contains the amplitude of the
time-dependent system in the basis state $|0\rangle $ ($|1\rangle $) after
algorithm step $i$.

Illustrating this with a concrete physical system, Fig.~1c shows a $2$%
-dimensional array of quantum dots (i.e., potential minima) with $4$ columns
of $2$ dots each. Column $i$ contains a state localized on the near dot ($%
c_{i,0}^{\dagger }\left\vert \mathrm{vac}\right\rangle $) and a state
localized on the far dot ($c_{i,1}^{\dagger }\left\vert \mathrm{vac}%
\right\rangle $); this sample algorithm has $N+1=4$ steps.

\begin{figure}[tbp]
\caption{Electron in array of quantum dots performing a single qubit, 4 step Deutsch-Jozsa algorithm for a function $f(x)$ where $f(0)=f(1)$.  Qubit starts in logical 0, undergoes a Hadamard gate, the mapping $\left|x\right> \rightarrow (-1)^{f(x)}\left|x\right>$, and a final Hadamard gate.  Each of the 4 columns of dots corresponds to an algorithm step. Near (far) dot of each column corresponds to logical 0 (1).  a) When AQC starts, Hamiltonian parameter $\protect\lambda =0$.  Ground state is localized on logical 0 dot of input column. b) As we increase $\protect\lambda$, potential of first columns increases and tunneling between columns turns on. Ground state spills from input step into later steps.  c) When $\protect \lambda =1$, ground state amplitudes in final column carry calculation results.}
\leavevmode
\includegraphics[height=7cm,angle=0]{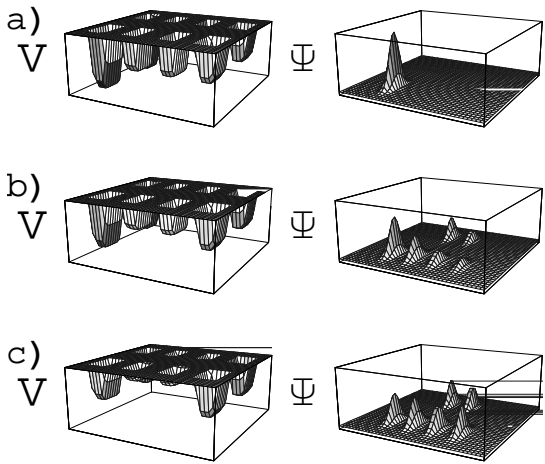}
\end{figure}

It is convenient to group creation operators into row vectors $%
C_{i}^{\dagger }\equiv \left[ c_{i,0}^{\dagger }\,\,\,\,c_{i,1}^{\dagger }%
\right] $. Then, the (unnormalized) ground state containing the results of
the algorithm is $\left\vert \Psi ^{N}\right\rangle =(C_{0}^{\dagger }\left[ 
{{1 \atop 0}}\right] +C_{1}^{\dagger }U_{1}\left[ {{1 \atop 0}}\right]
+\cdots +C_{N}^{\dagger }U_{N}\cdots U_{1}\left[ {{1 \atop 0}}\right]
)\left\vert \mathrm{vac}\right\rangle $. The results, stored in the states $%
c_{N,0}^{\dagger }\left\vert \mathrm{vac}\right\rangle $ and $%
c_{N,1}^{\dagger }\left\vert \mathrm{vac}\right\rangle $, can be extracted
reliably \cite{Mizel:04}. To execute the GSQC, one realizes a specific
Hamiltonian $H(1)$ whose ground state is the time-independent state we have
identified. When the computation involves a single qubit, the Hamiltonian
takes the form $H(1)=\sum_{i=1}^{N}h^{i}(U_{i})$ where 
\begin{equation}
h^{i}(U_{i})\equiv \mathcal{E}\left[ C_{i}^{\dagger }-C_{i-1}^{\dagger
}U_{i}^{\dagger }\right] \left[ C_{i}-U_{i}C_{i-1}\right] ,
\label{Hamsingle}
\end{equation}
and where $\mathcal{E}$ sets the energy scale.
Here we have used the quadratic form $C_{i}^{\dagger }VC_{j}=\sum_{x,y\in
\{0,1\}}v_{xy}c_{i,x}^{\dagger }c_{j,y}$, where $v_{xy}$ are the matrix
elements of $V$. In the quantum dots example of Fig.~1c, $H(1)$ controls the
onsite energy of each dot and the tunneling coupling between each dot in one
column and each dot in the next column. 
One can confirm $\left\vert \Psi ^{N}\right\rangle $ is an eigenstate with eigenvalue $0$ ($%
H(1)\left\vert \Psi ^{N}\right\rangle =0$); since every term
(\ref{Hamsingle}) is clearly positive semi-definite, $\left\vert \Psi
^{N}\right\rangle $ is the ground state \cite{grounddegennote}.

We have a time-independent state that contains the result of any given
standard algorithm for one qubit and is the ground state of a known
Hamiltonian $H(1)$. Now we show that placing the system in this ground state
can be done efficiently via the method of AQC, which constitutes a proof
that AQC\ can simulate the circuit model (for a single qubit). To do so we
introduce the Hamiltonian $H(s)=\sum_{i=1}^{N}h^{i}(\lambda (s)U_{i})$ where 
$\lambda :s\in \lbrack 0,1]\mapsto \lbrack 0,1]$ such that $\lambda (0)=0$
and $\lambda (1)=1$. If $\lambda =0$, then $h^{i}(0)=\mathcal{E}%
C_{i}^{\dagger }C_{i}$ reduces to a simple onsite energy term. There is then
no tunneling from algorithm step $i$ to algorithm step $i+1$. If $\lambda =1$%
, we recover $H(1)=\sum_{i=1}^{N}h^{i}(U_{i})$ with $h^{i}(U_{i})$ being the
full operator (\ref{Hamsingle}).

The (unnormalized) ground state of $H(s)$ is simply $\left\vert \Psi
^{N}\right\rangle $ as written above, but with a factor of $\lambda (s)$ in
front of each unitary operator $U_{i}$. Alternatively, the ground state is
given by the recursion relation 
\begin{equation}
\left\vert \Psi ^{j}(s)\right\rangle =(1+C_{j}^{\dagger }(\lambda
(s)U_{j})C_{j-1})\left\vert \Psi ^{j-1}(s)\right\rangle .  \label{recursion}
\end{equation}
Intuitively, the state of a $j$ step calculation, $\left\vert \Psi
^{j}(s)\right\rangle $, is formed by adding to the state of a $j-1$ step
calculation $\left\vert \Psi ^{j-1}(s)\right\rangle $ a term which
annihilates the particle at $j-1$ and creates a particle at $j$ with $%
\lambda (s)U_{j}$ applied to its state. The initial input state is
simply
$\left\vert \Psi ^{0}(s)\right\rangle =C_{0}^{\dagger }\left[ {1 \atop 0}
\right] \left\vert \mathrm{vac}\right\rangle $.

Fig.~1a shows how the wavefunction is localized on the input row when $%
\lambda =0$. As expected from the form of $h^{i}(\lambda(s)U_{i})$, as $%
\lambda $ increases, the onsite energy rises on rows $0,\ldots ,N-1$, as in
Fig.~1b. The tunneling matrix elements in (\ref{Hamsingle}) also begin to
turn on, and the ground state wavefunction starts to spill into the full
array of dots. When $\lambda$ reaches $1$, the wavefunction reaches the final row, as in Fig.~1c.

The traditional statement of the adiabatic theorem \cite{Schiff:68} is that
the increase in $\lambda (s)$ must be sufficiently gradual that the system
does not transition to an excited state as $s$ goes from $0$ to $1$. If the
time to take $s$ from 0 to 1 is $T$, transitions are suppressed if 
\mbox{$T\gg \hbar \,\,\,\, \mathrm{max}_{s}\left\vert \left\langle \chi
^{N}(s)\right\vert \frac{dH(s)}{ds}\left\vert \Psi ^{N}(s)\right\rangle
\right\vert/(E_{\chi }(s)-E_{\Psi }(s))^{2}
$}
where $\left\vert \chi ^{N}(s)\right\rangle $ is any excited eigenstate of $%
H(s)$, and $E_{\chi }(s)$ is its energy. Recent work has emphasized that
this is not necessarily the right condition, since what really matters is
not suppression of transitions throughout the entire adiabatic quantum
algorithm, but rather that the overlap between the ground state of $H(1)$
and the final adiabatic wavefunction be large \cite{Schaller:06,Jansen:06}.
An adiabatic condition arises of the form $T=\hbar O(\Delta E_{\min
}^{-1})$, where $\Delta E_{\min }$ is the minimum energy gap between
the ground and 
first excited state, as $s$ goes from $0$ to $1$
\cite{Schaller:06,Jansen:06}. Here we use this latter condition to
prove that AQC can simulate the circuit model efficiently; we omit the (similar but more complicated) proof
that uses the traditional adiabatic condition.
We note, however, that we will need to use our knowledge of the gap structure
(e.g., position of $\Delta E_{\min }$ as a function of $s$) in order to achieve the running time $T=\hbar O(\Delta E_{\min }^{-1})$, since it requires slowing the evolution down near $\Delta E_{\min }$.

To compute the minimum gap for an $N+1$ step calculation, we look for
solutions of $D_{N+1}^{2}(\bar{E})=0$, where $D_{N+1}^{2}\equiv \mathrm{det}%
\,\,(H(s)/\mathcal{E}-\bar{E})$, $H(s)$ is the $2(N+1)\times 2(N+1)$
Hamiltonian matrix, and $\bar{E}\equiv E/\mathcal{E}$. We first make a
unitary transformation to new operators $\tilde{C}_{i}\equiv (U_{i}^{\dagger
}\cdots U_{1}^{\dagger })C_{i}$ which transforms $H(s)=\sum_{i=1}^{N}h^{i}(%
\lambda (s)U_{i})$ to $\sum_{i=1}^{N}h^{i}(\lambda \mathcal{I})$, where $%
\mathcal{I}$ is the $2\times 2$ identity matrix. Writing out the matrix $H(s)$, we find the iterative
relation $D_{N+1}=(1+\lambda ^{2}-\bar{E})D_{N}-\lambda ^{2}D_{N-1}$.
The solutions to $D_{N+1}=0$ identify the exact
single qubit eigenenergies, which we find to be: $E_{0,s}=0$ and $E_{n,s}=\mathcal{E}((1-\lambda
(s))^{2}+2\lambda (s)(1-\cos \frac{\pi n}{N+1}))$ for $n=1,\ldots ,N$. By
minimizing the first excited state energy $E_{1,s}$ with respect to $\lambda 
$, one sees that $E_{1,s}\geq \mathcal{E}\sin ^{2}\frac{\pi }{(N+1)}=%
\mathcal{E}\,\,O(1/N^{2})$.  The minimum occurs when $\lambda = \cos \frac{\pi}{N+1}$.  Thus $\Delta E_{\min }=E_{1,s}-E_{0,s}=\mathcal{E}\,\,O(1/N^{2})$, and the simulation time is $T=\hbar O(\Delta E_{\min
}^{-1})=(\hbar /\mathcal{E})O(N^{2})$. Since this is polynomial we see that
AQC can efficiently simulate the circuit model.

\textit{Multiple qubits.}--- The recursion relation (\ref{recursion}) generalizes immediately to algorithms involving $M$ noninteracting qubits: $%
\left\vert \Psi ^{j}(s)\right\rangle =\Pi _{A=1}^{M}(1+C_{A,j}^{\dagger
}(\lambda U_{A,j})C_{A,j-1})\left\vert \Psi ^{j-1}(s)\right\rangle $ where $%
\left\vert \Psi ^{0}(s)\right\rangle =\Pi _{A=1}^{M}C_{A,0}^{\dagger }\left[ 
  {1 \atop 0}\right] \left\vert \mathrm{vac}\right\rangle $.
The multiple qubit Hamiltonian is just the sum of the single qubit
Hamiltonians $H(s)=\sum_{A=1}^{M}\sum_{i=1}^{N}h_{A}^{i}(\lambda
(s)U_{A,i})$; one can verify that $H(s)\left\vert \Psi
^{N}(s)\right\rangle =0$ for 
arbitrary $\lambda $. The AQC procedure for non-interacting qubits simply
consists of the single qubit procedure applied to each qubit independently.

Now, we allow the qubits to interact via two-qubit gates such as a
controlled-\textsc{NOT} (\textsc{CNOT}). Suppose the algorithm specifies a 
\textsc{CNOT} gate between qubits $A$ and $B$ at line $j$. Then, instead of
applying the factors $(I+C_{A,j}^{\dagger }(\lambda U_{A,j})C_{A,j-1})$ and $%
(I+C_{B,j}^{\dagger }(\lambda U_{B,j})C_{B,j-1})$ to $\left\vert \Psi
^{j-1}(s)\right\rangle $, we write 
\begin{eqnarray}
\lefteqn{\left\vert \Psi ^{j}(s)\right\rangle =(I+c_{A,j,0}^{\dagger
}\lambda c_{A,j-1,0}C_{B,j}^{\dagger }(\lambda I)C_{B,j-1}}  \notag \\
&&+c_{A,j,1}^{\dagger }\lambda c_{A,j-1,1}C_{B,j}^{\dagger }(\lambda \sigma
_{x})C_{B,j-1})\left\vert \Psi ^{j-1}(s)\right\rangle .  \notag
\end{eqnarray}%
If qubit $A$ is in state $0$, this operator applies $(I+C_{B,j}^{\dagger
}(\lambda I)C_{B,j-1})$ to qubit $B$. This is just the usual recursion
relation factor that subjects qubit $B$ to an \textsc{IDENTITY} gate. The
factor for a \textsc{NOT} gate, $(I+C_{B,j}^{\dagger }(\lambda \sigma
_{x})C_{B,j-1}))$, is applied to $B$ if $A$ is in state $1$.

When a \textsc{CNOT} gate is present, $H(s)$ needs to be changed so that we
still have $H(s)\left\vert \Psi ^{N}(s)\right\rangle =0$. One replaces terms 
$h_{A}^{j}(\lambda U_{A,j})$ and $h_{B}^{j}(\lambda U_{B,j})$ in $H(s)$ with 
\begin{equation}
h_{A,B}^{j}(\lambda ,{\mathrm{CNOT}})=h_{A,B}^{j}(\text{\textsc{\textrm{ID}}}%
)+h_{A,B}^{j}(\text{\textsc{\textrm{N}}})+h_{A,B}^{j}(\text{\textsc{\textrm{P}}}).
\label{hCNOT}
\end{equation}%
Here $h_{A,B}^{j}($\textsc{\textrm{ID}}$)=\mathcal{E}(C_{B,j}c_{A,j,0}-\lambda
^{2}C_{B,j-1}c_{A,j-1,0})^{\dag}\\\times (C_{B,j}c_{A,j,0}-\lambda
^{2}C_{B,j-1}c_{A,j-1,0})$, and $h_{A,B}^{j}($\textsc{\textrm{N}}$)=%
\mathcal{E}(C_{B,j}c_{A,j,1}-\lambda ^{2}\sigma _{x}C_{B,j-1}c_{A,j-1,1})^{\dag }\\ \times(C_{B,j}c_{A,j,1}-\lambda ^{2}\sigma _{x}C_{B,j-1}c_{A,j-1,1})$
are two-particle analogues of the one-particle \textsc{IDENTITY} gate $%
h_{A}^{j}(\lambda \mathcal{I})$ and \textsc{NOT} gate $h_{A}^{j}(\lambda
\sigma _{x})$ defined by (\ref{Hamsingle}). The third term $%
h_{A,B}^{j}({\mathrm{P}})=\mathcal{E}\sum_{i<j,k\geq j}C_{A,i}^{\dagger
}C_{A,i}C_{B,k}^{\dagger }C_{B,k}+C_{A,k}^{\dagger }C_{A,k}C_{B,i}^{\dagger
}C_{B,i}$ imposes an energy penalty on states in which one qubit has gone
through the CNOT gate without the other. Since $h_{A,B}^{j}(\lambda ,{%
\mathrm{CNOT}})$ is positive semi-definite along with all the other terms in 
$H(s)$, and as one verifies $H(s)\left\vert \Psi ^{N}(s)\right\rangle =0$, $%
\left\vert \Psi ^{N}(s)\right\rangle $ is still the ground state \cite%
{grounddegennote}.

To determine the effect of a \textsc{CNOT} gate on the gap, consider first a
simple calculation with $M=2$ qubits and a single \textsc{CNOT} at row $j$.
Divide the Hamiltonian into $H=H_{0}+H_{1}$, where $H_{1}$ is just the 
\textsc{CNOT} gate (\ref{hCNOT}) and $H_{0}$ contains all of the single
qubit terms. We know all of the exact eigenstates and eigenvalues of $H_{0}$
from the single qubit analysis of $D_{N+1}$ above. If the interaction $H_{1}$
were absent, the qubits would simply occupy these eigenstates of $H_{0}$
independently. Let $\left\vert Z\right\rangle $ be a two-qubit state
satisfying $H_{0}\left\vert Z\right\rangle =0$, where both qubits are in
zero-energy ground states. Let $\left\vert \bar{Z}\right\rangle $ be a state
in which at least one qubit is excited. Our single qubit analysis of $D_{N+1}
$ yields the exact result $\left\langle \bar{Z}\right\vert H_{0}\left\vert 
\bar{Z}\right\rangle \geq \mathcal{E}\sin ^{2}\frac{\pi }{(N+1)}=\mathcal{E}%
O(1/N^{2})$.

The \textsc{CNOT} Hamiltonian $H_{1}$ couples these eigenstates of $%
H_{0}$. It can be diagonalized analytically in the small basis of ground
states $\left\vert Z\right\rangle $; for all states $\left\vert
Z\right\rangle $ orthogonal to the computationally meaningful $\left\vert
\Psi ^{N}(s)\right\rangle $, we find $\left\langle Z\right\vert H\left\vert
Z\right\rangle =\left\langle Z\right\vert H_{1}\left\vert Z\right\rangle
= \mathcal{E}O(1/N^{2})$. This is a rigorous variational \emph{upper bound%
} of the true gap (because the small number of $|Z\rangle $ states are not a
basis for the full $[2(N+1)]^{2}$-dimensional Hilbert space), and it is also
a good estimate of the \emph{exact} energy gap, as confirmed in the $2$
qubit numerical calculation shown in Fig.~2. As a result, AQC requires a
time $T=(\hbar /\mathcal{E})O(N^{2})$. While this estimate is intuitively
correct and numerically verified,
what we need for our equivalence proof is a lower bound, which we now show
is $\mathcal{E}O(1/N^{4})$. To obtain this bound, note that an arbitrary
excited state of $H=H_{0}+H_{1}$ can be written as $\alpha \left\vert
Z\right\rangle +\beta \left\vert \bar{Z}\right\rangle $ for some unique
normalized $\left\vert Z\right\rangle $ and $\left\vert \bar{Z}\right\rangle 
$. Since $H_{1}$ is positive semi-definite, we find $\left\langle
Z|H_{1}|Z\right\rangle \left\langle \bar{Z}|H_{1}|\bar{Z}\right\rangle \geq
\left\vert \left\langle Z|H_{1}|\bar{Z}\right\rangle \right\vert ^{2}$.
Given this inequality, we find that the minimum of $\left\langle
H\right\rangle $ with respect to $\alpha $ and $\beta $ satisfies 
\begin{equation}
\left\langle H\right\rangle \geq \left\langle Z\right\vert H\left\vert
Z\right\rangle \left\langle \bar{Z}\right\vert H_{0}\left\vert \bar{Z}%
\right\rangle /(\left\langle Z\right\vert H\left\vert Z\right\rangle
+\left\langle \bar{Z}\right\vert H\left\vert \bar{Z}\right\rangle ).
\label{Hbound}
\end{equation}%
We can estimate the numerator using the $\mathcal{E}O(1/N^{2})$ bounds on $%
\left\langle \bar{Z}\right\vert H_{0}\left\vert \bar{Z}\right\rangle $ and $%
\left\langle Z\right\vert H\left\vert Z\right\rangle $ stated above. Since
the denominator of (\ref{Hbound}) certainly does not increase with $N$, the
energy $\left\langle H\right\rangle $ is at least $\mathcal{E}O(1/N^{4})$.

A similar argument works even when the system has many qubits and many 
\textsc{CNOT} gates. We write the many-qubit Hamiltonian as $H=H_{0}+H_{1}$,
where $H_{1}$ includes all of the \textsc{CNOT} gates (\ref{hCNOT}) and $%
H_{0}$ contains all of the single qubit terms. The exact eigenstates and
eigenenergies of $H_{0}$ are immediately known from the single qubit
analysis. An arbitrary state can be written $\alpha \left\vert
Z\right\rangle +\beta \left\vert \bar{Z}\right\rangle $ where in $\left\vert
Z\right\rangle $ all qubits are in ground states of $H_{0}$ while in $%
\left\vert \bar{Z}\right\rangle $ there is at least one excited qubit. We
still have $\left\langle \bar{Z}\right\vert H_{0}\left\vert \bar{Z}%
\right\rangle \geq \mathcal{E}\sin ^{2}\frac{\pi }{(N+1)}=\mathcal{E}%
O(1/N^{2})$. We can also show that $\left\langle Z\right\vert H\left\vert
Z\right\rangle \geq \mathcal{E}O(1/N^{2})$ since for each term in $%
\left\vert Z\right\rangle $ there is always at least one CNOT gate that
contributes to its energy. Using (\ref{Hbound}), we find the same $\mathcal{E%
}O(1/N^{4})$ bound on $\left\langle H\right\rangle $. To facilitate
extraction of the results of the computation, it is important that
when the system is measured every qubit have a large amplitude on the
final row $N$. As Ref.~\cite{Mizel:02} shows, it is straightforward to
adjust the Hamiltonian to ensure this happens.  However, the reduction
of qubit amplitude at earlier stages of the calculation leads to a
reduction of the gap.  The estimate/upper bound becomes
$\mathcal{E}O(1/N^{2}M)$ and the lower bound is
$\mathcal{E}O(1/N^{4}M^{2})$.

\begin{figure}[tbp]
\leavevmode
\includegraphics[height=5cm,angle=0]{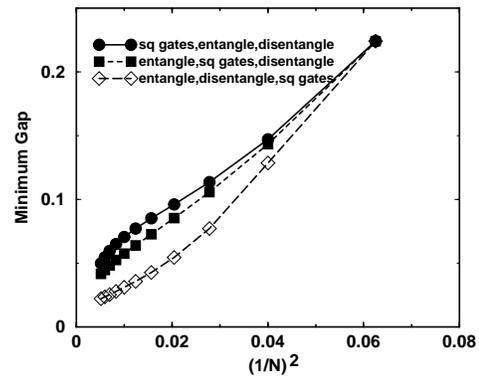}
\caption{Minimum gap in units of $\mathcal{E}$ for a 2 qubit system of $N$
steps. Results are shown for computations that entangle the two qubits into
a Bell state and then disentangle then. A string of single qubit gates is
also included at one of three stages of the computation. The minimum gap is
roughly a linear function of $1/N^2$.}
\end{figure}

We have presented an explicit adiabatic procedure that will carry a system
adiabatically into a ground state containing the result of an arbitrary
standard quantum computation. Ref.~\cite{Mizel:04} shows how to use quantum
teleportation to trade an arbitrary GSQC with $N$ steps and $M$
qubits for a different GSQC with $7$ steps, $(2N-1)M$ qubits. Once
the Hamiltonian is adjusted to facilitate extraction of the results \cite%
{Mizel:02}, the running time of the new calculation is
$T=(\hbar /\mathcal{E})O(N^{2}M^{2})$.  The upper bound/estimate
of the gap yields
$T=(\hbar /\mathcal{E})O(NM)$, which is proportional to the
\textquotedblleft volume\textquotedblright\ of the algorithm.

\begin{acknowledgments}
We thank M.L. Cohen for useful discussions. We gratefully acknowledge
support from the David and Lucile Packard Foundation, Research Innovation
Grant No. R10815, and NSF Grant No. PHY99-07949 (to A.M.), NSF Grant No.
CCF-0523675, and ARO-QA Grant No. W911NF-05-1-0440 (to D.A.L).
\end{acknowledgments}


\begin{thebibliography}{99}
\bibitem{comment} Note that in our formalism two particles never occupy the
same site, so the particle statistics is undetectable; we introduce
second-quantized notation for convenience.
\bibitem{grounddegennote}
Note that it is always necessary to introduce a perturbation to select $%
\left\vert \Psi ^{N}\right\rangle $ as the unique ground state over a state
that looks just like $\left\vert \Psi ^{N}\right\rangle $ but starts in $%
|1\rangle $ rather than $|0\rangle $. This corresponds to selecting the
input to the calculation to be $0$ rather than $1$. The perturbation we add
is $-\delta \mathcal{E}\,\,\sum_{A=1}^{M}c_{A,0,0}^{\dagger }c_{A,0,0}$, ,
where $\delta \mathcal{E}\ll \mathcal{E}$. This ensures a gap of $O(1/N)$
between $\left\vert \Psi ^{N}\right\rangle $ and states with undesired input
values, pushing them far above the low lying excited states of energy $\leq
O(1/N^{2})$. So, they play no role in determining the gap except in the
teleportation scheme discussed at the end of the paper. See \cite{Mizel:04}
for details.
\end{thebibliography}

\end{document}